\documentstyle [12pt, epsfig]{article}
\begin{document}

\hfill {CUMQ/HEP 126}

\hfill {\today}

\vskip 0.5in   \baselineskip 24pt

{\Large  \bigskip
      \centerline{\Large $\Delta M_K$ and $\epsilon_K$
                  in the left-right supersymmetric model} }

\vskip .6in
\def\bar{\overline}

\centerline{Mariana Frank \footnote{Email: mfrank@vax2.concordia.ca}
and Shuquan Nie \footnote{Email: sxnie@alcor.concordia.ca}}
\bigskip
\centerline {\it Department of Physics, Concordia University, 1455 De
Maisonneuve Blvd. W.}
\centerline {\it Montreal, Quebec, Canada, H3G 1M8}

\vskip 0.5in

{\narrower\narrower We perform a complete analysis of $\Delta S=2$
processes in the kaon system and evaluate $\Delta M_K$ and
$\epsilon_K$ in the left-right supersymmetric model. We include 
analytic expressions
for the contributions of gluinos, neutralinos and charginos. We
obtain general constraints on off-diagonal mass terms between the 
first two generations
of both down-type and up-type squarks. In the down-squark sector, we 
compare the results
with gluino-only estimates.  In the up-squark
sector, we find a complete set of bounds on all combinations of 
chirality conserving
or chirality flipping parameters. Finally, we comment on the size of 
the bounds obtained
by imposing left-right symmetry in the squark sector.}

      PACS number(s): 12.60.Jv, 13.25.Es, 14.40.Aq

\newpage

\section{Introduction}

Flavour changing neutral currents (FCNC) and charge-parity (CP)
violation are still
outstanding problems of electroweak theories. In the standard model
(SM), both are
understood to originate from the Cabibbo-Kobayashi-Maskawa (CKM) matrix, which
mixes quark flavours. The smallness of such phenomena is understood in terms
of the smallness of the off-diagonal elements of the CKM matrix. Extensions of
the SM, such as supersymmetry, introduce new FCNC and CP
parameters, which are {\it a priory} large. Restricting such parameters to fit
within the experimental results puts important constraints on the
supersymmetric
parameters and drastically reduces the parameter space.

Numerous analyses of such phenomena have been carried out in the context of the
minimal supersymmetric extension of the standard model (MSSM)
\cite{susy}. Some of
these analyses have some model-independent features, but the particle
content is MSSM. In a
series of recent papers, we investigated the constraints imposed on
the parameter
space of the left-right supersymmetric model (LRSUSY) from FCNC and CP
violation in the B system
\cite{fn}. The LRSUSY has gained considerable
interest in the literature for several interesting phenomenological
features, not least of
which is its ability to provide a solution to the SUSY-CP problem
\cite{mohras}. The
Hermitean structure of Yukawa matrices and soft-supersymmetry
breaking parameters above the right-handed breaking scale insures the
smallness of the electric dipole moments, a known problem in the MSSM. However,
off-diagonal elements of these parameters
are allowed to develop CP phases,  which play a role in B and K decay
parameters.

In this work we propose to extend the analysis of the flavor changing 
parameters
in the B system to K system,
in particular to the study of $\Delta
M_K$ and $\epsilon_K$. There are several important differences
between the B system and
K system. The K system is subject to many more hadronic
uncertainties than the B
system, in particular the mass difference $\Delta M_K$ is affected by long
distance contributions, which are difficult to estimate reliably. However,
improved QCD corrections in the calculation of
$K^0-{\bar K}^0$ mixing \cite{qcd}, NLO corrections to the $\Delta
S=2$ Hamiltonian
\cite{nlo}, and lattice calculations of the full set of parameters entering the
kaon mixing matrix elements \cite{lattice}, have raised the
theoretical accuracy of
predicting kaon parameters. Additionally, the kaon system would place
restrictions on different off-diagonal mass parameters and as such, provide
complimentary  information about flavor violation in the LRSUSY. The 
determination of
$\epsilon_K$ would provide a more complete picture on the constraints on CP
violation in this model.

Our paper is organized as follows: in section II we give a brief
description of the
model and sources of CP and flavor violation. Complete analytical results for
gluino, neutralino, gluino-neutralino and chargino contributions to
$\Delta M_K$
and $\epsilon_K$ are presented in section III, followed by a numerical analysis
in section IV.  We reach our conclusions in section V, and in the
appendix we give
a summary of chargino and neutralino mixing in the LRSUSY, as well as
the loop and vertex functions used, for self-sufficiency  of the paper.

\section{\bf Description of the LRSUSY Model}

The minimal left-right extension to the supersymmetric standard model
is based on the gauge group
$SU(3)_C \times SU(2)_L \times SU(2)_R \times U(1)_{B-L}$. The matter
fields of the model consist of three families of quark and lepton chiral
superfields with the following transformations under the gauge group
\begin{eqnarray}
Q&=&\left (\begin{array}{c}
u\\ d \end{array} \right ) \sim \left ( 3,2, 1, \frac13 \right ),~~
Q^c=\left (\begin{array}{c}
d^c\\u^c \end{array} \right ) \sim \left ( 3^{\ast},1, 2, -\frac13
\right ),\nonumber \\
L&=&\left (\begin{array}{c}
\nu\\ e\end{array}\right ) \sim\left ( 1,2, 1, -1 \right ),~~
L^c=\left (\begin{array}{c}
e^c \\ \nu^c \end{array}\right ) \sim \left ( 1,1, 2, 1 \right ),
\end{eqnarray}
where the numbers in the brackets denote the quantum numbers under
$SU(3)_C$,  $SU(2)_L$, $SU(2)_R$ and $U(1)_{B-L}$ respectively. The Higgs
sector consists of the bidoublet and triplet Higgs superfields
\begin{eqnarray}
\Phi_1 = \left (\begin{array}{cc}
\Phi^0_{11}&\Phi^+_{11}\\ \Phi_{12}^-& \Phi_{12}^0
\end{array}\right) \sim \left (1,2,2,0 \right),~~~~~~~~~~~~~
\Phi_2=\left (\begin{array}{cc}
\Phi^0_{21}&\Phi^+_{21}\\ \Phi_{22}^-& \Phi_{22}^0
\end{array}\right) \sim \left (1,2,2,0 \right),~~~~~~~ \nonumber \\
\Delta_{L}  = \left(\begin{array}{cc}
\frac {1}{\sqrt{2}}\Delta_L^-&\Delta_L^0\\
\Delta_{L}^{--}&-\frac{1}{\sqrt{2}}\Delta_L^-
\end{array}\right) \sim \left(1,3,1,-2\right), ~~\delta_{L}  =
\left(\begin{array}{cc}
\frac {1}{\sqrt{2}}\delta_L^+&\delta_L^{++}\\
\delta_{L}^{0}&-\frac{1}{\sqrt{2}}\delta_L^+
\end{array}\right) \sim \left(1,3,1,2\right),\nonumber \\
\Delta_{R}  =
\left(\begin{array}{cc}
\frac {1}{\sqrt{2}}\Delta_R^-&\Delta_R^0\\
\Delta_{R}^{--}&-\frac{1}{\sqrt{2}}\Delta_R^-
\end{array}\right) \sim \left(1,1,3,-2\right),~~ \delta_{R}  =
\left(\begin{array}{cc}
\frac {1}{\sqrt{2}}\delta_R^+&\delta_R^{++}\\
\delta_{R}^{0}&-\frac{1}{\sqrt{2}}\delta_R^+
\end{array}\right) \sim \left(1,1,3,2\right).
\end{eqnarray}
The bidoublet Higgs superfields appear in all LRSUSY and serve to
break the symmetry down to the $SU(2)_L \times U(1)_{Y}$ group. 
Supplementary Higgs
representations are needed to break $SU(2)_R$ symmetry spontaneously
at high $\Lambda_R$:
the choice of the adjoint representation is necessary to
support the
seesaw mechanism. Note that the number of fields in the Higgs sector is doubled
with respect to the non-supersymmetric version to ensure anomaly
cancellations. The
most general superpotential involving these superfields is
\begin{eqnarray}
\label{superpotential}
W & = & {\bf Y}_{Q}^{(i)} Q^T\Phi_{i}i \tau_{2}Q^{c} + {\bf Y}_{L}^{(i)}
L^T \Phi_{i}i \tau_{2}L^{c} + i({\bf Y}_{LR}L^T\tau_{2} \Delta_L L +
{\bf Y}_{LR}L^{cT}\tau_{2}
\Delta_R L^{c}) \nonumber \\
& & + \mu_{LR}\left [Tr (\Delta_L  \delta_L +\Delta_R
\delta_R)\right] + \mu_{ij}Tr(i\tau_{2}\Phi^{T}_{i} i\tau_{2} \Phi_{j})
+W_{NR},
\end{eqnarray}
where $W_{NR}$ denotes (possible) non-renormalizable terms arising
from higher scale
physics or Planck scale effects~\cite{recmohapatra}. These terms are
necessary for
R-parity conservation of the ground state, for the case in which
SUSY breaking scales are above $M_{W_{R}}$~\cite{km}. In addition, the
potential also includes soft supersymmetry breaking terms
\begin{eqnarray}
\label{eq:soft}
{\cal L}_{soft}&=&{\bf A}_{Q}^{i}{\bf Y}_{Q}^{(i)}{\tilde Q}^T\Phi_{i}
i\tau_{2}{\tilde Q}^{c}+ {\bf A}_{L}^{i}{\bf Y}_{L}^{(i)}{\tilde L}^T \Phi_{i}
i\tau_{2}{\tilde L}^{c} + i{\bf A}_{LR} {\bf Y}_{LR}({\tilde L}^T\tau_{2}
\Delta_L{\tilde  L} + L^{cT}\tau_{2} \Delta_R{\tilde L}^{c})
\nonumber \\
& & +{ m}_{\Phi}^{(ij) 2}
\Phi_i^{\dagger}  \Phi_j  + \left[( m_{L}^2)_{ij}{\tilde
l}_{Li}^{\dagger}{\tilde
l}_{Lj}+ (m_{R}^2)_{ij}{\tilde l}_{Ri}^{\dagger}{\tilde l}_{Rj}
\right] + \left[( m_{Q_L}^2)_{ij}{\tilde
Q}_{Li}^{\dagger}{\tilde
Q}_{Lj}+ (m_{Q_R}^2)_{ij}{\tilde Q}_{Ri}^{\dagger}{\tilde Q}_{Rj}
\right] \nonumber \\
& & - M_{LR}^2 \left[
Tr(  \Delta_R  \delta_R)+ Tr(  \Delta_L
     \delta_L) +h.c.\right]
-  [B \mu_{ij} \Phi_{i} \Phi_{j}+h.c.].
\end{eqnarray}
Additional flavor and CP violation is generated in the 
non-supersymmetric version of the
model through a right-handed version of the Cabibbo-Kobayashi-Maskawa matrix,
$K_R$, which is measurable in left-right models, but not in the SM.
Additional constraints can be imposed if this matrix is
different from the one
in the left-handed sector. However, here we are interested mainly in
the sources of flavor and CP breaking coming from the supersymmetric
sector, so we demand
manifest left-right symmetry $K_R=K_L$. Then the effects of the
right-handed currents are suppressed by the mass of the heavy $W_R$ boson.
Much less is
known about the right-handed supersymmetric partners of the gauge 
bosons and their
contributions to flavor and CP violation could be significant. In the
quark-squark sector, flavor violation arises mainly from the Yukawa 
couplings $Y_Q$ and the
trilinear scalar mixing parameters $A_Q$, through the squark mass matrices.

In what follows, we parameterize all the unknown soft breaking
parameters coming
mostly from the scalar mass matrices using the mass insertion approximation
\cite{MI}. We choose a basis for fermion and sfermion states
in which all the couplings of these particles to neutral gauginos are flavor
diagonal and parametrize flavor changes in
the non-diagonal squark propagators through mixing parameters
$(\delta^q_{ij})_{AB}$, where $i,j=1, 2, 3$ and $A,B=L,R$. The 
dimensionless flavor
mixing parameters used are
\begin{eqnarray}
\label{massins}
(\delta^{q}_{ij})_{LL}&=&\frac{(m^2_{\tilde{q},ij})_{LL}}{m_{\tilde q}^2},~~~
(\delta^{q}_{ij})_{RR}=\frac{(m^2_{\tilde{q},ij})_{RR}}{m_{\tilde
q}^2}, \nonumber \\
(\delta^{q}_{ij})_{LR}&=&\frac{(m^2_{\tilde{q},ij})_{LR}}{m_{\tilde q}^2},~~~
(\delta^{q}_{ij})_{RL}=\frac{(m^2_{\tilde{q},ij})_{RL}}{m_{\tilde q}^2},
\end{eqnarray}
where $m_{\tilde q}^2$ is the average squark mass and
$(m^2_{\tilde{q},ij})_{AB}$ are the
off-diagonal elements which mix squark flavors for both left- and
right- handed squarks with $q=u,d$.

\section{The analytic formulas}

\subsection{Effective Hamiltonian for the $\Delta S=2$ process in the LRSUSY}

The contributions of the left-right supersymmetric model to the
$\Delta S=2$ process
are given by the effective Hamiltonian
\begin{equation}
{\cal H}_{eff}^{\Delta S=2} =\sum_i [ C_i(\mu) Q_i(\mu) +
{\tilde C}_i(\mu) {\tilde Q}_i(\mu)].
\end{equation}
where the relevant operators entering the sum are
\begin{eqnarray}
Q_{1}&=&{\bar d}_L^\alpha \gamma_\mu s_L^\alpha {\bar
d}_L^\beta \gamma_\mu s_L^\beta,
\nonumber \\
{\tilde Q}_{1}&=&{\bar d}_R^\alpha \gamma_\mu s_R^\alpha {\bar
d}_R^\beta \gamma_\mu s_R^\beta,
\nonumber \\
Q_{2}&=&{\bar d}_L^\alpha s_R^\alpha {\bar
d}_L^\beta s_R^\beta,
\nonumber \\
{\tilde Q}_{2}&=&{\bar d}_R^\alpha s_L^\alpha {\bar
d}_R^\beta s_L^\beta,
\nonumber \\
Q_{3}&=&{\bar d}_L^\alpha  s_R^\beta {\bar
d}_L^\beta  s_R^\alpha,
\nonumber \\
{\tilde Q}_3&=&{\bar d}_R^\alpha  s_L^\beta {\bar
d}_R^\beta  s_L^\alpha,
\nonumber \\
Q_{4}&=&{\bar d}_L^\alpha s_R^\alpha {\bar
d}_R^\beta s_L^\beta,
\nonumber \\
Q_{5}&=&{\bar d}_L^\alpha  s_R^\beta {\bar
d}_R^\beta  s_L^\alpha,
\end{eqnarray}
where $q_{R,L}=P_{R,L} q$ with $P_{R,L} =(1 \pm \gamma_5)/2$, and
$\alpha,~ \beta$ are color indices.
The Wilson coefficients $C_{i}$ and ${\tilde C}_i$ are initially evaluated
at the electroweak or soft supersymmetry breaking scale, then evolved down
to the scale $\mu$. Because of the left-right symmetry,
we must consider all contributions from both chirality operators.

\begin{figure}
\centerline{ \epsfysize 3.0in
\rotatebox{360}{\epsfbox{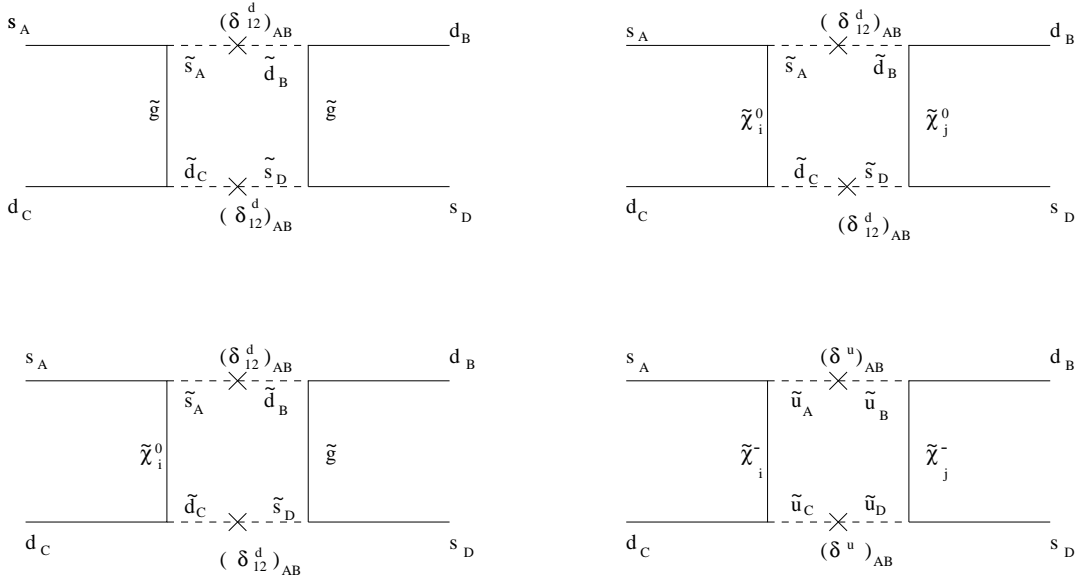}}  }
\caption{ Leading supersymmetric box diagrams contributing to
$K^0-\bar{K}^0$ mixing, $A, ~B, ~C, ~D=(L,~R)$.
Crossed diagrams (not shown) are also included in the calculation. }
\protect \label{feynman}
\end{figure}

The $K^0-{\bar K}^0$ mixing is mediated through the box diagrams shown in Fig.
\ref{feynman}.
Below we give a comprehensive list of all box diagram contributions for
$C_i$ and $\tilde {C}_i$. The notations of various vertices, mixing
matrices and functions
are defined in the appendix.

\begin{equation}
C_i=C_i^{\tilde{g}}+C_i^{\tilde{g}\tilde{\chi}^0}+C_i^{\tilde{\chi}^0}+C_i^{\tilde{\chi}^-},
\end{equation}
with
\begin{eqnarray}
C_1^{\tilde{g}}&=& -\frac{\alpha_s^2}{ m^2_{\tilde{q}}} x_{\tilde {q}
\tilde {g}}
\left ( \frac{1}{9} F(x_{\tilde{q} \tilde{g}}, 1)
-\frac{11}{36} G(x_{\tilde{q} \tilde{g}}, 1)\right ) (\delta_{12}^d)^{2}_{LL},
\nonumber \\
C_2^{\tilde{g}}&=& -\frac{\alpha_s^2}{ m^2_{\tilde{q}}} x_{\tilde {q}
\tilde {g}}
\frac{17}{18} F(x_{\tilde{q} \tilde{g}}, 1) (\delta_{12}^d)^{2}_{RL},
\nonumber \\
C_3^{\tilde{g}}&=& \frac{\alpha_s^2}{6 m^2_{\tilde{q}}} x_{\tilde {q}
\tilde {g}}  F(x_{\tilde{q} \tilde{g}}, 1) (\delta_{12}^d)^{2}_{RL},
\nonumber \\
C_4^{\tilde{g}}&=& -\frac{\alpha_s^2}{ m^2_{\tilde{q}}} x_{\tilde {q}
\tilde {g}}  \left[ \left ( \frac{7}{3} F(x_{\tilde{q} \tilde{g}}, 1)
+\frac{1}{3} G(x_{\tilde{q} \tilde{g}}, 1)\right )
(\delta_{12}^d)_{LL}(\delta_{12}^d)_{RR}\right.
+ \left. \frac{11}{18} G(x_{\tilde{q} \tilde{g}}, 1)
(\delta_{12}^d)_{LR}(\delta_{12}^d)_{RL}\right],
\nonumber \\
C_5^{\tilde{g}}&=& \frac{\alpha_s^2}{ m^2_{\tilde{q}}} x_{\tilde {q}
\tilde {g}} \left[ \left (\frac{5}{9} G(x_{\tilde{q} \tilde{g}}, 1)
-\frac{1}{9} F(x_{\tilde{q} \tilde{g}}, 1) \right )
(\delta_{12}^d)_{LL}(\delta_{12}^d)_{RR}\right.
-\left. \frac {5}{6} G(x_{\tilde{q} \tilde{g}}, 1)
(\delta_{12}^d)_{LR}(\delta_{12}^d)_{RL}\right],
\nonumber \\
C_1^{\tilde g \tilde{\chi}^0}&=& - \frac{\alpha_s \alpha_W}{3 m_{\tilde q}^2}
x_{\tilde {q} \tilde {g}} \sum_{i=1}^9
\left[ G_{DL}^{i} G_{DL}^{\star i}
G(x_{\tilde{q} \tilde{g}}, x_{\tilde{\chi}_i^0 \tilde{g}}) \right.
\nonumber \\
& & -  \left.   (G_{DL}^{\star i} G_{DL}^{\star i}
+ G_{DL}^{i} G_{DL}^{i})
\sqrt{x_{\tilde{\chi}_i^0 \tilde{g}}} F(x_{\tilde{q} \tilde{g}},
x_{\tilde{\chi}_i^0 \tilde{g}}) \right](\delta_{12}^d)^{2}_{LL},
\nonumber \\
C_2^{\tilde g \tilde{\chi}^0}&=& -\frac{\alpha_s \alpha_W}{3 m_{\tilde q}^2}
x_{\tilde {q} \tilde {g}} \sum_{i=1}^9
     G_{DL}^{\star i} G_{DL}^{\star i}
\sqrt{x_{\tilde{\chi}_i^0 \tilde{g}}} F(x_{\tilde{q} \tilde{g}},
x_{\tilde{\chi}_i^0 \tilde{g}})(\delta_{12}^d)^{2}_{RL},
\nonumber \\
C_3^{\tilde g \tilde{\chi}^0}&=& -\frac{\alpha_s \alpha_W}{3 m_{\tilde q}^2}
x_{\tilde {q} \tilde {g}} \sum_{i=1}^9
     G_{DL}^{\star i} G_{DL}^{\star i}
\sqrt{x_{\tilde{\chi}_i^0 \tilde{g}}}
F(x_{\tilde{q} \tilde{g}},x_{\tilde{\chi}_i^0 \tilde{g}})
(\delta_{12}^d)^{2}_{RL},
\nonumber \\
C_4^{\tilde g \tilde{\chi}^0}&=& \frac{\alpha_s \alpha_W}{ m_{\tilde q}^2}
x_{\tilde {q} \tilde {g}} \sum_{i=1}^9
     \left[ \left( G_{DR}^{i} G_{DL}^{ i}+G_{DR}^{\star i}
G_{DL}^{\star i}\right)
G(x_{\tilde{q} \tilde{g}}, x_{\tilde{\chi}_i^0 \tilde{g}})\right.
\nonumber \\
 &   &-   \left.   2( G_{DL}^{ i} G_{DR}^{\star i}+ G_{DL}^{\star i} G_{DR}^{i})
\sqrt{x_{\tilde{\chi}_i^0 \tilde{g}}}
F(x_{\tilde{q} \tilde{g}},
x_{\tilde{\chi}_i^0 \tilde{g}})(\delta_{12}^d)_{LL}(\delta_{12}^d)_{RR}\right.
\nonumber \\
&  & -   \left. \left( G_{DL}^{ i} G_{DL}^{\star i}
+ G_{DR}^{ i} G_{DR}^{\star i}\right)
G(x_{\tilde{q} \tilde{g}}, x_{\tilde{\chi}_i^0 \tilde{g}})
(\delta_{12}^d)_{RL}(\delta_{12}^d)_{LR}\right],
\nonumber \\
C_5^{\tilde g \tilde{\chi}^0}&=& \frac{\alpha_s \alpha_W}{3 m_{\tilde q}^2}
x_{\tilde {q} \tilde {g}} \sum_{i=1}^9
     \left \{\left[ 2\left ( G_{DL}^{i} G_{DR}^{\star i}+G_{DL}^{\star
i}G_{DR}^{i}\right )\sqrt{x_{\tilde{\chi}_i^0 \tilde{g}}}
F(x_{\tilde{q} \tilde{g}},x_{\tilde{\chi}_i^0 \tilde{g}}) \right. \right.
\nonumber \\
& &-  \left. \left. \left ( G_{DR}^{i} G_{DL}^{i}+G_{DR}^{\star i}G_{DL}^{\star
i}\right )  G(x_{\tilde{q} \tilde{g}}, x_{\tilde{\chi}_i^0 \tilde{g}})
(\delta_{12}^d)_{LL}(\delta_{12}^d)_{RR} \right] +
(G_{DL}^{ i} G_{DL}^{\star i}+ G_{DR}^{i} G_{DR}^{\star i})\right.
\nonumber \\
&  & \times   \left.
\sqrt{x_{\tilde{\chi}_i^0 \tilde{g}}} F(x_{\tilde{q} \tilde{g}},
x_{\tilde{\chi}_i^0 \tilde{g}}) -(G_{DR}^{ i} G_{DR}^{\star i}
+ G_{DL}^{i} G_{DL}^{\star i}) G(x_{\tilde{q} \tilde{g}},
x_{\tilde{\chi}_i^0 \tilde{g}})
](\delta_{12}^d)_{LR}(\delta_{12}^d)_{RL} \right\},
\nonumber  \\
C_1^{\tilde{\chi}^0}&=& \frac{\alpha_W^2}{2 m_{\tilde{q}}^2}
\sum_{i,j=1}^9 x_{\tilde {q} \tilde{\chi}_i^0}
\left[ G_{DL}^{j} G_{DL}^{\star j} G_{DL}^{i} G_{DL}^{\star j}
G(x_{\tilde{q} \tilde{\chi}_i^0},
x_{\tilde{\chi}_j^0 \tilde{\chi}_i^0}) \right.
\nonumber \\
& & -  \left. 2G_{DL}^{j} G_{DL}^{\star i} G_{DL}^{j} G_{DL}^{\star i}
\sqrt{x_{\tilde{\chi}_j^0 \tilde{\chi}_i^0}}
F(x_{\tilde{q} \tilde{\chi}_i^0},
x_{\tilde{\chi}_j^0 \tilde{\chi}_i^0})\right](\delta_{12}^d)^{2}_{LL},
\nonumber \\
C_2^{\tilde{\chi}^0}&=& -\frac{\alpha_W^2}{m_{\tilde{q}}^2} \sum_{i,j=1}^9
x_{\tilde {q} \tilde{\chi}_i^0} G_{DR}^{j} G_{DR}^{j} G_{DL}^{\star i}
G_{DL}^{\star i} \sqrt{x_{\tilde{\chi}_j^0 \tilde{\chi}_i^0}}
     F(x_{\tilde{q} \tilde{\chi}_i^0},
x_{\tilde{\chi}_j^0 \tilde{\chi}_i^0})](\delta_{12}^d)^{2}_{RL},
\nonumber \\
C_3^{\tilde{\chi}^0}&=& -\frac{\alpha_W^2}{2 m_{\tilde{q}}^2} \sum_{i,j=1}^9
x_{\tilde {q} \tilde{\chi}_i^0} \left( G_{DR}^{j}G_{DR}^{j} G_{DL}^{\star i}
G_{DL}^{\star i} \right.
\nonumber \\
& & - \left. G_{DR}^{j}G_{DR}^{i} G_{DL}^{\star j}
G_{DL}^{\star i}\right) \sqrt{x_{\tilde{\chi}_j^0 \tilde{\chi}_i^0}}
F(x_{\tilde{q} \tilde{\chi}_i^0},
x_{\tilde{\chi}_j^0 \tilde{\chi}_i^0})(\delta_{12}^d)^{2}_{RL},
\nonumber \\
C_4^{\tilde{\chi}^0}&=& -\frac{\alpha_W^2}{2 m_{\tilde{q}}^2}
\sum_{i,j=1}^9 x_{\tilde {q} \tilde{\chi}_i^0}
G_{DR}^{j} G_{DL}^{\star j} G_{DL}^{i} G_{DR}^{\star i}
G(x_{\tilde{q} \tilde{\chi}_i^0},
x_{\tilde{\chi}_j^0 \tilde{\chi}_i^0})(\delta_{12}^d)_{RL}(\delta_{12}^d)_{LR},
\nonumber \\
C_5^{\tilde{\chi}^0}&=& \frac{2\alpha_W^2} {m_{\tilde{q}}^2} \sum_{i,j=1}^9
x_{\tilde {q} \tilde{\chi}_i^0}\left \{ \left[ G_{DR}^{i} G_{DL}^{i}
G_{DR}^{\star j} G_{DL}^{\star j} G(x_{\tilde{q} \tilde{\chi}_i^0},
x_{\tilde{\chi}_j^0 \tilde{\chi}_i^0})\right. \right.
\nonumber \\
& &  - \left. \left. 2 G_{DR}^{i} G_{DL}^{\star i} G_{DL}^{
j} G_{DR}^{\star j} \sqrt{x_{\tilde{\chi}_j^0 \tilde{\chi}_i^0}}
     F(x_{\tilde{q} \tilde{\chi}_i^0},
x_{\tilde{\chi}_j^0 \tilde{\chi}_i^0}) \right](\delta_{12}^d)_{LL}
(\delta_{12}^d)_{RR}\right.
\nonumber \\
&  &  -  \left. \left[G_{DL}^{i} G_{DR}^{i} G_{DR}^{\star j}
G_{DL}^{\star j} G(x_{\tilde{q} \tilde{\chi}_i^0},
x_{\tilde{\chi}_j^0 \tilde{\chi}_i^0})\right. \right.
\nonumber \\
& &- \left. \left. 2 G_{DL}^{i} G_{DR}^{ j} G_{DL}^{\star
j} G_{DR}^{\star i} \sqrt{x_{\tilde{\chi}_j^0 \tilde{\chi}_i^0}}
F(x_{\tilde{q} \tilde{\chi}_i^0},
x_{\tilde{\chi}_j^0 \tilde{\chi}_i^0})\right] \right\}(\delta_{12}^d)_{LR}
(\delta_{12}^d)_{RL},
\nonumber \\
C_1^{\tilde{\chi}^-} &=&- \frac{\alpha_W^2}{6m_{\tilde{q}}^2}
\left( \sum_{h,k=1}^3 K^{\ast}_{h2} (\delta^u_{hk})_{LL}
K_{k1} \right)^2
\sum_{i,j=1}^{4} x_{\tilde {q} \tilde{\chi}_i^-}
G_{UL}^{j} G_{UL}^{\star i} G_{UL}^{i} G_{UL}^{\star j}
G(x_{\tilde{q} \tilde{\chi}_i^-},
x_{\tilde{\chi}_j^- \tilde{\chi}_i^-}),
\nonumber  \\
C_2^{\tilde{\chi}^-} &=& 0,
\nonumber \\
C_3^{\tilde{\chi}^-} &=& \frac{\alpha_W^2}{m_{\tilde{q}}^2}
\left( \sum_{h,k=1}^3 K^{\ast}_{h2} (\delta^u_{hk})_{RL}
K_{k1} \right)^2 \sum_{i,j=1}^{4}  x_{\tilde {q} \tilde{\chi}_i^-}
G_{UR}^{j} G_{UL}^{\star i} G_{UR}^{i}G_{UL}^{\star j}
   \sqrt{x_{\tilde{\chi}_j^- \tilde{\chi}_i^-}}
F(x_{\tilde{q} \tilde{\chi}_i^-},
x_{\tilde{\chi}_j^- \tilde{\chi}_i^-}),
\nonumber  \\
C_4^{\tilde{\chi}^-} &=&\frac{\alpha_W^2}{12m_{\tilde{q}}^2}
\sum_{h,k,m,n=1}^3 K^{\ast}_{h2} (\delta^u_{hk})_{RL} K_{k1}
K^{\ast}_{m2} (\delta^u_{mn})_{LR} K_{n1}
\nonumber \\
&  & \times 
\sum_{i,j=1}^{4} x_{\tilde {q} \tilde{\chi}_i^-}
G_{UR}^{j} G_{UL}^{ i}G_{UL}^{\star i}G_{UR}^{\star j}
G(x_{\tilde{q} \tilde{\chi}_i^-},
x_{\tilde{\chi}_j^- \tilde{\chi}_i^-}),
\nonumber  \\
C_5^{\tilde{\chi}^-} &=&- \frac{\alpha_W^2}{m_{\tilde{q}}^2}
\sum_{h,k,m,n=1}^3 K^{\ast}_{h2} (\delta^u_{hk})_{LL} K_{k1}
K^{\ast}_{m2} (\delta^u_{mn})_{RR} K_{n1}
\nonumber \\
&  & \times   \sum_{i,j=1}^{4} x_{\tilde {q} \tilde{\chi}_i^-}
G_{UR}^{i} G_{UL}^{j}
G_{UR}^{\star j} G_{UL}^{\star i}
\sqrt{x_{\tilde{\chi}_j^- \tilde{\chi}_i^-}}
F(x_{\tilde{q} \tilde{\chi}_i^-},
x_{\tilde{\chi}_j^- \tilde{\chi}_i^-}),
\end{eqnarray}
where $x_{ab}=m^2_a/m^2_b$. In the above expressions we have neglected the
contributions coming from higgsino vertices because of proportionality
to the masses of the first two generations of quarks.
The coefficients $\tilde {C}_i$ are obtained from
the coefficients $C_i$
\begin{eqnarray}
{\tilde C}_i^{\tilde{g},~(\tilde{g} \tilde \chi^0,~\tilde
\chi^0,~\tilde \chi^-)}&=&
C_i^{\tilde{g},~(\tilde{g} \tilde \chi^0,~\tilde \chi^0,~\tilde \chi^-)} ( L
\leftrightarrow R ).
\end{eqnarray}

\subsection{Hadronic Matrix Elements}
The matrix elements of the operators $Q_i$ between $K^0-\bar{K}^0$ mesons in
the vacuum insertion approximation (VIA) \cite{VIA} are given by
\begin{eqnarray}
\left < K^0| Q_1 | \bar{K}^0 \right >_{VIA} &=& \frac{1}{3} m_K f_K^2,
\nonumber \\
\left < K^0| Q_2 | \bar{K}^0 \right >_{VIA} &=& -\frac{5}{24} \left (
\frac{m_K}{m_s+m_d} \right )^2
m_K f_K^2,
\nonumber \\
\left < K^0| Q_3 | \bar{K}^0 \right >_{VIA} &=& \frac{1}{24} \left (
\frac{m_K}{m_s+m_d} \right )^2
m_K f_K^2,\nonumber \\
\left < K^0| Q_4 | \bar{K}^0 \right >_{VIA} &=&  \left
[\frac{1}{24}+\frac{1}{4} \left ( \frac{m_K}{m_s+m_d} \right )^2
\right ]
m_K f_K^2,\nonumber \\
\left < K^0| Q_5 | \bar{K}^0 \right >_{VIA} &=& \left [
\frac{1}{8}+\frac{1}{12} \left ( \frac{m_K}{m_s+m_d} \right )^2
\right ]
m_K f_K^2,
\end{eqnarray}
where $m_K=497.67$ MeV, $m_s$ and $m_d$ are the masses of the $K^0$
meson, s and d quark respectively,
$f_K=159.8$ MeV is the decay constant of $K^0$ mesons. The
expressions for $\tilde{Q}_{1-3}$ are same as those of $Q_{1-3}$.

To take into account nonperturbative effects, the B parameters, which 
are scale dependent,
are defined as
\begin{eqnarray}
\left < K^0| Q_1 | \bar{K}^0 \right >_{VIA} &=& \frac{1}{3} m_K f_K^2 B_1(\mu),
\nonumber \\
\left < K^0| Q_2 | \bar{K}^0 \right >_{VIA} &=& -\frac{5}{24} \left (
\frac{m_K}{m_s+m_d} \right )^2
m_K f_K^2 B_2(\mu),
\nonumber \\
\left < K^0| Q_3 | \bar{K}^0 \right >_{VIA} &=& \frac{1}{24} \left (
\frac{m_K}{m_s+m_d} \right )^2
m_K f_K^2 B_3(\mu),\nonumber \\
\left < K^0| Q_4 | \bar{K}^0 \right >_{VIA} &=&  \frac{1}{4} \left (
\frac{m_K}{m_s+m_d} \right )^2
m_K f_K^2 B_4(\mu),\nonumber \\
\left < K^0| Q_5 | \bar{K}^0 \right >_{VIA} &=& \frac{1}{12} \left (
\frac{m_K}{m_s+m_d} \right )^2
m_K f_K^2 B_5(\mu),
\end{eqnarray}
with the numerical values at $\mu=2$ GeV
\begin{eqnarray}
m_s(\mu)=125 ~\mathrm{MeV}, & & m_d(\mu)=7 ~\mathrm{MeV},
\nonumber \\
B_1(\mu)=0.60(6), & & B_2(\mu)=0.66(4),
     \nonumber \\
B_3(\mu)=1.05(12), & & B_4(\mu)=1.03(6),
\nonumber \\
B_5(\mu)=0.73(10).
\end{eqnarray}
The coefficients at the scale of $\mu$ are given by
\begin{equation}
C_r (\mu)=\sum_i \sum_s (b_i^{(r,s)}+ \eta c_i^{(r,s)} ) \eta^{a_i} C_s (M),
\end{equation}
where $\eta=\alpha_s(M)/\alpha_s(m_t)$ and we have chosen
$M=(m_{\tilde{g}}+m_{\tilde{q}})/2$.
The numerical coefficients $a_i, b_i^{(r,s)}, c_i^{(r,s)}$ can be found in Ref.
\cite{ciuchini}.

Putting all the above together, we can calculate the mass difference
$\Delta M_K$ and
CP violating parameter $\epsilon_K$.
\begin{eqnarray}
\Delta M_K &=& 2 \mathrm{Re} \left < K^0 |
{\cal H}_{eff}^{\Delta S=2} | \bar{K}^0 \right >,
\nonumber \\
\epsilon_K &=& \frac{1}{\sqrt{2} \Delta M_K} \mathrm{Im} \left < K^0 |
{\cal H}_{eff}^{\Delta S=2} | \bar{K}^0 \right >.
\end{eqnarray}
By setting the calculated expressions to their experimental results, we obtain
restrictions on the sources of flavor and CP violation in the LRSUSY.

\section{Numerical Analysis}

In this section we present the results of our analysis for the
individual bounds on both
$(\delta^d_{12})_{AB}$ and $(\delta^u_{12})_{AB}$, obtained by
selecting only one
source of flavor violation and neglecting interference between
different sources. In
retrospect, the hierarchical structure of the flavor-violating
parameters in the kaon system disfavors accidental cancellations.

We obtain the constraints from $\Delta M_K$ by requiring that
the LRSUSY contribution proportional to a
single $\delta$
parameter does not exceed the present experimental value. The SM
contribution is
proportional to the $cs$ and $cd$ elements of the $K$ matrix and thus
independent of SUSY
contributions. A common assumption is that the SM and supersymmetric
contributions simply add, and several bounds are obtained in the
literature with this simplification. We consider only
supersymmetric effects here, and short-distance effects only, so that, if
all the other effects are additive, the constraints we obtain
are more conservative than previous analyses. This assumption also avoids
introducing extra
parameters, such as the relative phase of various contributions and mass
parameters for the Higgs bosons into the calculations.

In the case of $\epsilon_K$, the constraints are also obtained requiring
the LRSUSY contribution alone to saturate the experimental value for
$\epsilon_K$. The SM
contribution depends on the phase in the CKM matrix, which is
another free
parameter, but the SM result for $\epsilon_K$ is always positive.
Thus, setting the CKM
phase to zero gives conservative bounds on the supersymmetric flavor violation
parameters.

We analyze cases in which the supersymmetric
partners have masses around the weak scale, so we will assume relatively light
superpartner masses.
We choose all trilinear scalar couplings in the soft supersymmetry
breaking Lagrangian
to be universal: $A_{ij}=A \delta_{ij}$ and
$\mu_{ij}=\mu_H \delta_{ij}$, and we fix $A$ to be $100$ GeV and the 
Higgsino mixing parameter $\mu_H=200$ GeV
throughout the analysis.

To constrain $(\delta^d_{12})_{AB}$, we include contributions from gluino,
neutralino and gluino-neutralino graphs. Because of the large number 
of neutralinos in the
LRSUSY, the last two contributions are what distinguishes this model
from the MSSM. We set $M_L=M_R=500$ GeV, and allow the
squark and gluino masses to vary.  We obtain bounds on
Re$(\delta^d_{12})_{AB}(\delta^d_{12})_{CD}$ with
$AB, CD=LL, RR, LR, RL$ from $\Delta M_K$. The bounds obtained are 
presented in Table {\bf 1}. We include in brackets, for self-check and comparison, the 
bounds  obtained by including
only gluino contributions. The results are in agreement with previous 
evaluations available in the literature.
Specifically, compared with Ref.
\cite{ciuchini}, the bounds on $(\delta_{12}^d)_{LL}$ are slightly 
weaker, mainly due to
that we set the SM contribution zero, while Ref.  \cite{ciuchini} 
subtracted the SM
contribution (to the coefficient $C_1$).
The results share the same general features as the bounds
obtained in the MSSM \cite{ciuchini}, that the restrictions on the
Re$(\delta)$ parameters become stronger with decreasing
$x=m_{\tilde{g}}^2/m_{\tilde{q}}^2$. There is, to a good
approximation, left-right symmetry in the flavor-violating parameters, and the
chirality flipping
parameters are more restricted than the chirality conserving
parameters. As expected
from the MSSM evaluations, the gluino diagram dominates. Neutralino 
contributions are suppressed
by $(\alpha_W/\alpha_s)^2$ compared with gluino contributions,  thus 
they are negligible; while gluino-neutralino
contributions are suppressed by $\alpha_W/\alpha_s$ and, from Table 
{\bf 1}, they contribute at
most 10 \% for heavy gluino and light neutralino masses.
  These features persist for bounds on
$\mathrm{Im}(\delta^d_{12})_{AB}(\delta^d_{12})_{CD}$ coming from
$\epsilon_K$, presented in Tables {\bf 2}. We include there too, in 
brackets, the gluino-only contributions.

We present next restrictions on
$(\delta^u_{12})_{AB}(\delta^u_{12})_{CD}$ coming from
the chargino sector. It is there that we expect to encounter the
most significant  deviations from
the MSSM, since the LRSUSY includes a right-handed gaugino. This 
analysis allows
restrictions of a
full complement of up-type squark flavor violating mass insertions,
similar to the ones
obtained for the down-type squark sector. In Table {\bf 3}, we present the
bounds obtained on
the real parts of the mass insertions, under the assumption that only
one insertion
dominates. The results obtained are almost same as the corresponding
ones in the MSSM \cite{chargino}, where only bounds on $\delta 
^u_{LL}$ were given.
All constraints become weaker with increasing squark mass
$m_{\tilde q}$. The bounds appear to be approximately left-right symmetric.
$\mathrm{Im}(\delta^u_{12})_{AB}$ shown in Table {\bf 4} share the 
general features
for $\mathrm{Re}(\delta^u_{12})_{AB}$, except that
there the bounds are stronger by about a factor of 10.
The bounds on $\mathrm{Im}(\delta^u_{12})_{LL}$
are in agreement with the bounds found in Ref. \cite{chargino}.
We expect these to be further
strengthened by an analysis of
$\epsilon^{\prime}/{\epsilon}$ \cite{eprimee}.

In LR models, one cannot generally assume that there is only one type 
of mass insertion at a time.
The bounds obtained in the above considerations, by allowing only one 
non-zero mass insertion at a time, show approximate
left-right symmetry. Thus it is worthwhile to consider possible 
bounds which respect the left-right symmetry.
If we let $\delta_{LR}^{d,u}=\delta_{RL}^{d,u}$ and 
$\delta_{LL}^{d,u}=\delta_{RR}^{d,u}$, the parameters reduce to
$\mathrm{Re~ (Im)}$$(\delta_{12}^{d,u})_{LR}$ and  $\mathrm{Re~ 
(Im)}$$(\delta_{12}^{d,u})_{LL}$.
Effectively, this means that we are allowing two mass insertions at 
the same time. The bounds are shown in Table {\bf
5}-{\bf 8}.  They are much more restrictive than the ones obtained by 
allowing only one source of flavor violation
because several contributions from different sources are added 
together.  For both $\delta_{LL}^{d,u}$ and
$\delta_{LR}^{d,u}$, the bounds follow from the most restrictive 
bounds from the previous tables, coming from
$\sqrt{|\mathrm{Re(Im)} (\delta^{u,d}_{12})_{LL} 
(\delta^{u,d}_{12})_{RR} |} $ and $\sqrt{|\mathrm{Re (Im)} (
\delta^{u,d}_{12})_{LR} (\delta^{u,d}_{12})_{RL} |} $. These bounds 
are purely a consequence of left-right symmetry and
have no counterpart in other models one could compare with.

\section{Conclusions}

We have performed a complete analysis of the $K^0-{\bar K}^0$ mixing
in a general left-right supersymmetric model, where flavor violation
is parametrized in
a model-independent way using the mass insertion approximation. We
have provided general
upper bounds on mass insertions for either the up or down squark
sectors, under the
assumption that the supersymmetric contributions alone do not
exceed the experimental value of $\Delta M_K$. This analysis provides bounds on
$\mathrm{Re}(\delta^d_{12})_{AB}$ and $\mathrm{Re}(\delta^u_{12})_{AB},~A,B=L,R$.
Similarly, for $\mathrm{Im}(\delta^d_{12})_{AB}$ and
$\mathrm{Im}(\delta^u_{12})_{AB}$, we obtain
upper bounds by requiring that the supersymmetric contributions
do not exceed the experimental values for $\epsilon_K$, {\it i.e.} setting
the CKM phase to
zero. The bounds for the down-squark sector are obtained including
gluino-neutralino and neutralino diagrams (in addition to gluino
only) and are similar to those coming from the MSSM.
In the up-squark sector, the bounds are
much weaker than in
the neutral sector, justifying the assumption of gluino domination of
this process. The bounds on $\mathrm{Re}(\delta^u_{12})_{LL}$ and
$\mathrm{Im}(\delta^u_{12})_{LL}$ are comparable to the ones
obtained in the MSSM. In this sector, we also obtain, for the first 
time, bounds on the
real and imaginary parts of all combinations of chirality conserving 
or flipping parameters.
As a general feature, the imaginary parts are
bound to be smaller
than their real counterparts by a factor of 10. If we impose 
left-right symmetry in the squark mass matrix,
$\delta_{LR}^{d,u}=\delta_{RL}^{d,u}$ and 
$\delta_{LL}^{d,u}=\delta_{RR}^{d,u}$, the bounds are dramatically 
different
and better than bounds on single mass insertion by one or more orders 
of magnitude. A realisation of such
a scenario would be a promising sign for the manifestation of 
left-right supersymmetry.

\vskip0.2in

\noindent {\bf Acknowledgements}

This work was funded in part by NSERC of Canada (SAP0105354).

\newpage

\begin{appendix}

\noindent {\Large {\bf Appendix}}
\label{appendix}

For self-sufficiency,  we list the mass-squared
matrices for charginos and neutralinos,
relevant Feynman rules and functions used for this calculation.

The terms relevant to the masses of charginos in the Lagrangian are
\begin{equation}
{\cal L}_C=-\frac{1}{2}(\psi^+, \psi^-) \left ( \begin{array}{cc}
                                                          0 & X^T \\
                                                          X & 0
                                                        \end{array}
                                                \right ) \left (
\begin{array}{c}
                                                                 \psi^+ \\
                                                                 \psi^-
                                                                 \end{array}
                                                          \right ) + H.c. \ ,
\end{equation}
where $\psi^+=(-i \lambda^+_L, -i \lambda^+_R, \tilde{\phi}_{u1}^+,
\tilde{\phi}_{d1}^+)^T$
and $\psi^-=(-i \lambda^-_L, -i \lambda^-_R, \tilde{\phi}_{u2}^-,
\tilde{\phi}_{d2}^-)^T$, and
\begin{equation}
X=\left( \begin{array}{cccc}
                              M_L & 0 & g_L \kappa_u & 0  \\
                              0 & M_R & g_R \kappa_u & 0  \\
                              0 & 0 & 0 & -\mu_H  \\
                              g_L \kappa_d & g_R \kappa_d & -\mu_H & 0
                 \end{array}
           \right ),
\end{equation}
where we have taken, for simplification, $\mu_{ij}=\mu_H \delta_{ij}$. The chargino mass
eigenstates $\chi_i$ are obtained by
\begin{eqnarray}
\tilde{\chi}_i^+=V_{ij}\psi_j^+, \ \tilde{\chi}_i^-=U_{ij}\psi_j^-, \
i,j=1, \ldots 4,
\end{eqnarray}
with $V$ and $U$ unitary matrices satisfying
\begin{equation}
U^* X V^{-1} = M_D,
\end{equation}
The diagonalizing matrices $U^*$ and $V$ are obtained by
computing the eigenvectors corresponding
to the eigenvalues of $X^{\dagger} X$ and $X X^{\dagger}$, respectively.

The terms relevant to the masses of neutralinos in the Lagrangian are
\begin{equation}
{\cal L}_N=-\frac{1}{2} {\psi^0}^T Y \psi^0  + H.c. \ ,
\end{equation}
where $\psi^0=(-i \lambda_L^3, -i \lambda_R^3, -i \lambda_V,
\tilde{\phi}_{u1}^0, \tilde{\phi}^0_{u2},
\tilde{\phi}_{d1}^0, \tilde{\phi}^0_{d2}, \tilde{\Delta}_R^0,
\tilde{\delta}_R^0 )^T $,
and
\begin{equation}
Y=\left( \begin{array}{ccccccccc}
                 M_L & 0 & 0 & \frac{g_L \kappa_u}{2} & 0 & 0 & -
\frac{g_L \kappa_d}{2} & 0 & 0 \\
                 0 & M_R & 0 & \frac{g_R \kappa_u}{2} & 0 & 0 &
-\frac{g_R \kappa_d}{2} & - g_R v_R \sin \phi & -g_R v_R \cos \phi \\
                 0 & 0 & M_V & 0 & 0 & 0 & 0 & 2 g_V v_R \sin \phi & 2
g_V v_R \cos \phi \\
                 \frac{g_L \kappa_u}{2} & \frac{g_R \kappa_u}{2}
&
0 & 0 & 0 & 0 & -\mu_H & 0 & 0  \\
                 0 & 0 & 0 & 0 & 0 & -\mu_H & 0 & 0 & 0 \\
                 0 & 0 & 0 & 0 & -\mu_H & 0 & 0 & 0 & 0 \\
                 -\frac{g_L \kappa_d}{2} & -\frac{g_R
\kappa_d}{2}
& 0 & -\mu_H & 0 & 0 & 0 & 0 & 0  \\
                 0 & -g_R v_R \sin \phi & 2 g_V v_R \sin \phi & 0 & 0 &
0 & 0 & 0 &
-\mu_H \\
                 0 & -g_R v_R \cos \phi & 2 g_V v_R \cos \phi & 0 & 0 &
0 & 0 & -\mu_H & 0
                 \end{array}
           \right ),
\end{equation}
where $\sin \phi =\tan \theta_W$. The mass eigenstates are defined by
\begin{equation}
\tilde{\chi}^0_i=N_{ij} \psi^0_j \ (i,j=1,2, \ldots 9),
\end{equation}
where $N$ is a unitary matrix chosen such that
\begin{equation}
N^* Y N^{-1} = N_D,
\label{equationN}
\end{equation}
and $N_D$ is a diagonal matrix with non-negative entries.

The vertices of neutralino-quark-squark and chargino-quark-squark
given by the mixing martices $G_{DL,DR}$ and $G_{UL,UR}$ respectively, where
\begin{eqnarray}
G^{j}_{DL} &=& \left[\sin \theta_W Q_d N^{\prime}_{j1} + \frac{1}{\cos
\theta_W} (T^3_{d}-Q_d \sin^2 \theta_W)
        N^{\prime}_{j2} \right.
\nonumber \\
& -&\left. \frac{\sqrt{\cos 2 \theta_W }}{\cos \theta_W}
\frac{Q_u+Q_d}{2} N^{\prime}_{j3}\right]  \\
G^{j}_{DR} &=& -\left[\sin \theta_W Q_d N^{\prime}_{j1} - \frac{Q_d \sin^2
\theta_W}{\cos \theta_W}
        N^{\prime}_{j2} \right.
\nonumber \\
&+& \left. \frac{\sqrt{\cos 2 \theta_W }}{\cos \theta_W}
(T^3_{d}-Q_d \sin^2 \theta_W) N^{\prime}_{j3}\right] ,
     \\
G^{j}_{UL} &=& V_{j1}^{\ast} ,
     \\
G^{j}_{UR} &=& U_{j2}.
\end{eqnarray}

The relevant two-variable functions from the box diagrams are
\begin{equation}
\left( F(x,y), ~ G(x, y) \right)=x^2 \partial_a \partial_b \left( 
F^{\prime} (a,b,y), ~G^{\prime} (a,b,y) \right)
|_{a=b=x},
\end{equation}
with
\begin{eqnarray}
F^{\prime}(x,y,z)&=&-\frac{1}{x-y}\left \{ \frac{1}{x-z} \left[ 
\frac{x}{x-1} \log x
- (x \rightarrow z)\right]-(x \rightarrow y) \right \},
\\
G^{\prime}(x,y,z)&=&\frac{1}{x-y} \left \{ \frac{1}{x-z} \left[ 
\frac{x^2}{x-1} \log x
-\frac{3}{2}x - (x \rightarrow z)\right]-(x \rightarrow y)\right \}.
\end{eqnarray}

\end{appendix}

\bibliographystyle{unsrt}

\newpage

\begin{tabular}{|c|c|c|c|}  \hline
    & $m_{\tilde{q}}=200$ GeV &$m_{\tilde{q}}=500$ GeV &
$m_{\tilde{q}}=1000$ GeV \\
    \hline
    $x$ & \multicolumn{3}{c|}{$\sqrt{|\mathrm{Re}
(\delta^{d}_{12})_{LL}^{2}|} $} \\
\hline
0.25& $1.1 ~[1.1]\times 10^{-2}$

     & $3.1~[3.1]\times 10^{-2}$

     & $6.8~[6.8]\times 10^{-2}$

\\
1.0& $2.3~[2.6]\times 10^{-2}$

    & $6.3~[7.5]\times 10^{-2}$

    & $1.4~[1.6]\times 10^{-1}$

\\
4.0& $1.8~[0.66]\times 10^{-1}$

    & $2.1~[1.8]\times 10^{-1}$

    & $3.9~[4.0]\times 10^{-1}$

\\
\hline
    $x$ & \multicolumn{3}{c|}{$\sqrt{|\mathrm{Re}
(\delta^{d}_{12})_{LR}^{2}|} $} \\
\hline
0.25& $1.4~[1.4]\times 10^{-3}$

     & $3.6~[3.6]\times 10^{-3}$

     & $7.6~[7.7]\times 10^{-3}$

\\
1.0& $1.6~[1.5]\times 10^{-3}$

    & $4.2~[4.1]\times 10^{-3}$

    & $8.9~[8.7]\times 10^{-3}$

\\
4.0& $2.2~[2.3]\times 10^{-3}$

    & $6.0~[6.0]\times 10^{-3}$

    & $1.2~[1.3]\times 10^{-2}$

\\
\hline
    $x$ & \multicolumn{3}{c|}{$\sqrt{|\mathrm{Re}
(\delta^{d}_{12})_{RL}^{2}|} $} \\
    \hline
0.25& $1.4~[1.4]\times 10^{-3}$

     & $3.7~[3.7]\times 10^{-3}$

     & $7.7~[7.9]\times 10^{-3}$

\\
1.0& $1.5~[1.6]\times 10^{-3}$

    & $4.0~[4.2]\times 10^{-3}$

    & $8.4~[8.9]\times 10^{-3}$

\\
4.0& $2.3~[2.3]\times 10^{-3}$

    & $6.1~[6.2]\times 10^{-3}$

    & $1.3~[1.3]\times 10^{-3}$

\\
\hline
    $x$ & \multicolumn{3}{c|}{$\sqrt{|\mathrm{Re}
(\delta^{d}_{12})_{RR}^{2}|} $} \\
\hline
0.25& $1.1~[1.1]\times 10^{-2}$

     & $3.1~[3.1]\times 10^{-2}$

     & $6.8~[6.9]\times 10^{-2}$

\\
1.0& $2.5~[2.6]\times 10^{-2}$

    & $6.9~[7.5]\times 10^{-2}$

    & $1.5~[1.6]\times 10^{-1}$

\\
4.0& $8.9~[6.6]\times 10^{-2}$

    & $5.6~[1.8]\times 10^{-1}$

    & $7.9~[4.0]\times 10^{-1}$

\\
\hline
    $x$ & \multicolumn{3}{c|}{$\sqrt{|\mathrm{Re}  (
\delta^{d}_{12})_{LL} (\delta^{d}_{12})_{RR} |} $} \\
    \hline
0.25& $4.6~[4.2]\times 10^{-4}$

     & $1.2~[1.1]\times 10^{-3}$

     & $2.5~[2.3]\times 10^{-3}$

\\
1.0& $5.4~[5.2]\times 10^{-4}$

    & $1.4~[1.4]\times 10^{-3}$

    & $2.9~[2.8]\times 10^{-3}$

\\
4.0& $7.9~[7.8]\times 10^{-4}$

    & $2.0~[2.0]\times 10^{-3}$

    & $4.1~[4.2]\times 10^{-3}$
\\
\hline
    $x$ & \multicolumn{3}{c|}{$\sqrt{|\mathrm{Re}  (
\delta^{d}_{12})_{LR} (\delta^{d}_{12})_{RL} |} $} \\
    \hline
0.25& $6.3~[6.3]\times 10^{-4}$

     & $1.6~[1.6]\times 10^{-3}$

     & $3.2~[3.4]\times 10^{-3}$

\\
1.0& $1.1~[1.1]\times 10^{-3}$

    & $2.8~[2.9]\times 10^{-3}$

    & $5.4~[6.0]\times 10^{-3}$

\\
4.0& $2.5~[2.6]\times 10^{-3}$

    & $6.1~[6.8]\times 10^{-3}$

    & $1.1~[1.4]\times 10^{-2}$
\\
\hline
\end{tabular}

\vspace{0.5cm}
${\bf  Table~~ 1} ~~$ Limits on $\mathrm{Re} (\delta_{12}^d)_{AB}
(\delta_{12}^d)_{CD}$,
with A, B, C, D=(L,R) for different values of
$x=m_{\tilde{g}}^2/m_{\tilde{q}}^2$ with one mass insertion each time,
including gluino and neutralino
contributions and the SM contribution set to zero. The results in brackets
represent gluino-only contribution.

\begin{tabular}{|c|c|c|c|}  \hline
    & $m_{\tilde{q}}=200$ GeV &$m_{\tilde{q}}=500$ GeV &
$m_{\tilde{q}}=1000$ GeV \\
    \hline
    $x$ & \multicolumn{3}{c|}{$\sqrt{|\mathrm{Im}
(\delta^{d}_{12})_{LL}^{2}|} $} \\
\hline
0.25& $8.7~[8.8]\times 10^{-4}$

     & $2.5~[2.5]\times 10^{-3}$

     & $5.4~[5.5]\times 10^{-3}$

\\
1.0& $1.9~[2.1]\times 10^{-3}$

    & $5.1~[6.0]\times 10^{-3}$

    & $1.1~[1.3]\times 10^{-2}$

\\
4.0& $1.4~[0.53]\times 10^{-2}$

    & $1.7~[1.5]\times 10^{-2}$

    & $3.2~[3.2]\times 10^{-2}$

\\
\hline
    $x$ & \multicolumn{3}{c|}{$\sqrt{|\mathrm{Im}
(\delta^{d}_{12})_{LR}^{2}|} $} \\
\hline
0.25& $1.1~[1.1]\times 10^{-4}$

     & $2.9~[2.9]\times 10^{-4}$

     & $6.1~[6.2]\times 10^{-4}$

\\
1.0& $1.3~[1.2]\times 10^{-4}$

    & $3.4~[3.3]\times 10^{-4}$

    & $7.1~[7.0]\times 10^{-4}$

\\
4.0& $1.8~[1.8]\times 10^{-4}$

    & $4.8~[4.8]\times 10^{-4}$

    & $1.0~[1.0]\times 10^{-3}$

\\
\hline
    $x$ & \multicolumn{3}{c|}{$\sqrt{|\mathrm{Im}
(\delta^{d}_{12})_{RL}^{2}|} $} \\
    \hline
0.25& $1.1~[1.1]\times 10^{-4}$

     & $3.0~[3.0]\times 10^{-4}$

     & $6.2~[6.3]\times 10^{-4}$

\\
1.0& $1.2~[1.3]\times 10^{-4}$

    & $3.2~[3.4]\times 10^{-4}$

    & $6.7~[7.2]\times 10^{-4}$

\\
4.0& $1.8~[1.9]\times 10^{-4}$

    & $4.9~[5.0]\times 10^{-4}$

    & $1.0~[1.0]\times 10^{-3}$

\\
\hline
    $x$ & \multicolumn{3}{c|}{$\sqrt{|\mathrm{Im}
(\delta^{d}_{12})_{RR}^{2}|} $} \\
\hline
0.25& $8.8~[8.8]\times 10^{-4}$

     & $2.5~[2.5]\times 10^{-3}$

     & $5.4~[5.5]\times 10^{-3}$

\\
1.0& $2.0~[2.1]\times 10^{-3}$

    & $5.5~[6.0]\times 10^{-3}$

    & $1.2~[1.3]\times 10^{-2}$

\\
4.0& $7.2~[5.3]\times 10^{-3}$

    & $4.5~[1.5]\times 10^{-2}$

    & $6.3~[3.2]\times 10^{-2}$

\\
\hline
    $x$ & \multicolumn{3}{c|}{$\sqrt{|\mathrm{Im}  (
\delta^{d}_{12})_{LL} (\delta^{d}_{12})_{RR} |} $} \\
    \hline
0.25& $3.7~[3.4]\times 10^{-5}$

     & $9.7~[8.8]\times 10^{-5}$

     & $2.0~[1.8]\times 10^{-4}$

\\
1.0& $4.3~[4.2]\times 10^{-5}$

    & $1.1~[1.1]\times 10^{-4}$

    & $2.3~[2.2]\times 10^{-4}$

\\
4.0& $6.3~[6.2]\times 10^{-5}$

    & $1.6~[1.6]\times 10^{-4}$

    & $3.3~[3.3]\times 10^{-4}$
\\
\hline
    $x$ & \multicolumn{3}{c|}{$\sqrt{|\mathrm{Im}  (
\delta^{d}_{12})_{LR} (\delta^{d}_{12})_{RL} |} $} \\
    \hline
0.25& $5.0~[5.1]\times 10^{-5}$

     & $1.3~[1.3]\times 10^{-4}$

     & $2.5~[2.7]\times 10^{-4}$

\\
1.0& $8.9~[9.0]\times 10^{-5}$

    & $2.2~[2.3]\times 10^{-4}$

    & $4.3~[4.8]\times 10^{-4}$

\\
4.0& $2.0~[2.1]\times 10^{-4}$

    & $4.9~[5.4]\times 10^{-4}$

    & $9.2~[11.0]\times 10^{-4}$
\\
\hline
\end{tabular}

\vspace{0.5cm}
${\bf  Table~~ 2} ~~$ Limits on $\mathrm{Im} (\delta_{12}^d)_{AB}
(\delta_{12}^d)_{CD}$,
with A, B, C, D=(L,R) for different values of
$x=m_{\tilde{g}}^2/m_{\tilde{q}}^2$ with one mass insertion each time,
including gluino and neutralino
contributions and the SM contribution set to zero. The results in brackets
represent gluino-only contribution.

\begin{tabular}{|c|c|c|c|}  \hline
    $M_L $ $\slash$ $ m_{\tilde{q}}$ (GeV) & 300  & 500 & 700 \\
\hline
        & \multicolumn{3}{c|}{$\sqrt{|\mathrm{Re}
(\delta^{u}_{12})_{LL}^{2}|} $}
   \\
\hline
     150 &$6.2 \times 10^{-2}$

         &$8.1 \times 10^{-2}$

         &$1.0 \times 10^{-1}$\\
     250 &$9.0 \times 10^{-2}$

         &$1.0 \times 10^{-1}$

         &$1.2 \times 10^{-1}$ \\
     350 &$1.2 \times 10^{-1}$

         &$1.3 \times 10^{-1}$

         &$1.5 \times 10^{-1}$ \\
     450 &$1.7 \times 10^{-1}$

         &$1.6 \times 10^{-1}$

         &$1.7 \times 10^{-1}$ \\
\hline
         & \multicolumn{3}{c|}{$\sqrt{|\mathrm{Re}
(\delta^{u}_{12})_{RR}^{2}|} $}
   \\
\hline
     150 &$6.9 \times 10^{-2}$

         &$8.7 \times 10^{-2}$

         &$1.1 \times 10^{-1}$\\
     250 &$9.9 \times 10^{-2}$

         &$1.1 \times 10^{-1}$

         &$1.3 \times 10^{-1}$ \\
     350 &$1.4 \times 10^{-1}$

         &$1.4 \times 10^{-1}$

         &$1.5 \times 10^{-1}$ \\
     450 &$1.8 \times 10^{-1}$

         &$1.7 \times 10^{-1}$

         &$1.8 \times 10^{-1}$ \\
\hline
         & \multicolumn{3}{c|}{$\sqrt{|\mathrm{Re}
(\delta^{u}_{12})_{LR}^{2}|}, ~~~(\delta^u_{12})_{RL}=(\delta^u_{12})_{LR}$}
   \\
\hline
     150 &$1.6 \times 10^{-1}$

         &$3.0 \times 10^{-1}$

         &$5.0 \times 10^{-1}$\\
     250 &$1.3 \times 10^{-2}$

         &$2.0 \times 10^{-2}$

         &$2.9 \times 10^{-2}$ \\
     350 &$1.6 \times 10^{-2}$

         &$2.3 \times 10^{-2}$

         &$3.0 \times 10^{-2}$ \\
     450 &$1.7 \times 10^{-2}$

         &$2.2 \times 10^{-2}$

         &$2.9 \times 10^{-2}$ \\
\hline
        & \multicolumn{3}{c|}{$\sqrt{|\mathrm{Re}  (
\delta^{u}_{12})_{LL} (\delta^{u}_{12})_{RR} |} $} \\
\hline
     150 &$8.1 \times 10^{-2}$

         &$1.6 \times 10^{-1}$

         &$2.6 \times 10^{-1}$ \\
     250 &$6.9 \times 10^{-3}$

         &$1.0 \times 10^{-2}$

         &$1.5 \times 10^{-2}$ \\
     350 &$8.4 \times 10^{-3}$

         &$1.2 \times 10^{-2}$

         &$1.5 \times 10^{-2}$ \\
     450 &$8.8 \times 10^{-3}$

         &$1.1 \times 10^{-2}$

         &$1.5 \times 10^{-2}$ \\
\hline
         & \multicolumn{3}{c|}{$\sqrt{|\mathrm{Re}  (
\delta^{u}_{12})_{LR} (\delta^{u}_{12})_{RL} |} $} \\
\hline
     150 &$9.7 \times 10^{-3}$

         &$1.2 \times 10^{-2}$

         &$1.5 \times 10^{-2}$ \\
     250 &$1.4 \times 10^{-2}$

         &$1.5 \times 10^{-2}$

         &$1.8 \times 10^{-2}$ \\
     350 &$1.9 \times 10^{-2}$

         &$1.9 \times 10^{-2}$

         &$2.1 \times 10^{-2}$ \\
     450 &$2.5 \times 10^{-2}$

         &$2.4 \times 10^{-2}$

         &$2.5 \times 10^{-2}$ \\
\hline
\end{tabular}

\vspace{0.5cm}
${\bf  Table ~~3} ~~$ Limits on $\mathrm{Re} (\delta_{12}^u)_{AB}
(\delta_{12}^u)_{CD}$,
with A, B, C, D=(L,R) for different values of
$M_L=M_R$ and $m_{\tilde{q}}$ with one mass insertion each time,
with $\tan \beta=5$ and $\mu_H=200$ GeV.
The bounds  are insensitive to $\tan \beta$ in the range of $2-30$
and to $\mu_H$ in the range of $200-500$ GeV.

\begin{tabular}{|c|c|c|c|}  \hline
    $M_L $ $\slash$ $ m_{\tilde{q}}$ (GeV) & 300  & 500 & 700 \\
\hline
        & \multicolumn{3}{c|}{$\sqrt{|\mathrm{Im}
(\delta^{u}_{12})_{LL}^{2}|} $}
   \\
\hline
     150 &$5.0 \times 10^{-3}$

         &$6.5 \times 10^{-3}$

         &$8.2 \times 10^{-3}$\\
     250 &$7.2 \times 10^{-3}$

         &$8.4 \times 10^{-3}$

         &$9.9 \times 10^{-3}$\\
     350 &$1.0 \times 10^{-2}$

         &$1.1 \times 10^{-2}$

         &$1.2 \times 10^{-2}$\\
     450 &$1.3 \times 10^{-2}$

         &$1.3 \times 10^{-2}$

         &$1.4 \times 10^{-2}$\\
\hline
         & \multicolumn{3}{c|}{$\sqrt{|\mathrm{Im}
(\delta^{u}_{12})_{RR}^{2}|} $}
   \\
\hline
     150 &$5.5 \times 10^{-3}$

         &$7.0 \times 10^{-3}$

         &$8.7 \times 10^{-3}$\\
     250 &$7.9 \times 10^{-3}$

         &$8.9 \times 10^{-3}$

         &$1.0 \times 10^{-2}$ \\
     350 &$1.1 \times 10^{-2}$

         &$1.1 \times 10^{-2}$

         &$1.2 \times 10^{-2}$ \\
     450 &$1.4 \times 10^{-2}$

         &$1.4 \times 10^{-2}$

         &$1.5 \times 10^{-2}$ \\
\hline
         & \multicolumn{3}{c|}{$\sqrt{|\mathrm{Im}
(\delta^{u}_{12})_{LR}^{2}|},~~~(\delta^u_{12})_{RL}=(\delta^u_{12})_{LR} $}
   \\
\hline
     150 &$1.3 \times 10^{-2}$

         &$2.4 \times 10^{-2}$

         &$4.1 \times 10^{-2}$\\
     250 &$1.1 \times 10^{-3}$

         &$1.6 \times 10^{-3}$

         &$2.3 \times 10^{-3}$ \\
     350 &$1.3 \times 10^{-3}$

         &$1.8 \times 10^{-3}$

         &$2.4 \times 10^{-3}$ \\
     450 &$1.4 \times 10^{-3}$

         &$1.8 \times 10^{-3}$

         &$2.3 \times 10^{-3}$ \\
\hline
        & \multicolumn{3}{c|}{$\sqrt{|\mathrm{Im}  (
\delta^{u}_{12})_{LL} (\delta^{u}_{12})_{RR} |} $} \\
\hline
     150 &$6.5 \times 10^{-3}$

         &$1.2 \times 10^{-2}$

         &$2.1 \times 10^{-2}$\\
     250 &$5.5 \times 10^{-4}$

         &$8.4 \times 10^{-4}$

         &$1.2 \times 10^{-3}$\\
     350 &$6.8 \times 10^{-4}$

         &$9.3 \times 10^{-4}$

         &$1.2 \times 10^{-3}$\\
     450 &$7.1 \times 10^{-4}$

         &$9.2 \times 10^{-4}$

         &$1.2 \times 10^{-3}$ \\
\hline
         & \multicolumn{3}{c|}{$\sqrt{|\mathrm{Im}  (
\delta^{u}_{12})_{LR} (\delta^{u}_{12})_{RL} |} $} \\
\hline
     150 &$7.8 \times 10^{-4}$

         &$9.7 \times 10^{-4}$

         &$1.2 \times 10^{-3}$ \\
     250 &$1.1 \times 10^{-3}$

         &$1.2 \times 10^{-3}$

         &$1.4 \times 10^{-3}$ \\
     350 &$1.5 \times 10^{-3}$

         &$1.6 \times 10^{-3}$

         &$1.7 \times 10^{-3}$ \\
     450 &$2.0 \times 10^{-3}$

         &$1.9 \times 10^{-3}$

         &$2.0 \times 10^{-3}$ \\
\hline
\end{tabular}

\vspace{0.5cm}
${\bf  Table ~~4} ~~$ Limits on $\mathrm{Im} (\delta_{12}^u)_{AB}
(\delta_{12}^u)_{CD}$,
with A, B, C, D=(L,R) for different values of
$M_L=M_R$ and $m_{\tilde{q}}$ with one mass insertion each time,
with $\tan \beta=5$ and $\mu_H=200$ GeV. The bounds
are insensitive to $\tan \beta$ in the range of $2-30$ and to $\mu_H$
in the range of $200-500$ GeV.

\newpage

\begin{tabular}{|c|c|c|c|}  \hline
    & $m_{\tilde{q}}=200$ GeV &$m_{\tilde{q}}=500$ GeV &
$m_{\tilde{q}}=1000$ GeV \\
    \hline
    $x$ & \multicolumn{3}{c|}{$\sqrt{|\mathrm{Re}
(\delta^{d}_{12})_{LL}^{2}|},~~~ 
(\delta^{d}_{12})_{RR}=(\delta^{d}_{12})_{LL}$} \\
\hline
0.25& $4.7\times 10^{-4}$

     & $1.2\times 10^{-3}$

     & $2.5\times 10^{-3}$

\\
1.0& $5.4\times 10^{-4}$

    & $1.4\times 10^{-3}$

    & $2.9\times 10^{-3}$

\\
4.0& $7.9\times 10^{-4}$

    & $2.0\times 10^{-3}$

    & $4.1\times 10^{-3}$

\\
\hline
    $x$ & \multicolumn{3}{c|}{$\sqrt{|\mathrm{Re}
(\delta^{d}_{12})_{LR}^{2}|},~~~(\delta^{d}_{12})_{RL}=(\delta^{d}_{12})_{LR} 
$} \\
\hline
0.25& $8.2\times 10^{-4}$

     & $2.0\times 10^{-3}$

     & $3.9\times 10^{-3}$

\\
1.0& $6.9\times 10^{-3}$

    & $8.8\times 10^{-3}$

    & $1.2\times 10^{-2}$

\\
4.0& $2.1\times 10^{-3}$

    & $5.9\times 10^{-3}$

    & $1.4\times 10^{-2}$

\\
\hline
\end{tabular}

\vspace{0.5cm}
${\bf  Table~~ 5} ~~$ Limits on $\mathrm{Re} (\delta_{12}^d)_{AB}
(\delta_{12}^d)_{CD}$,
with A, B, C, D=(L,R) for different values of
$x=m_{\tilde{g}}^2/m_{\tilde{q}}^2$ with two mass insertions each time,
including gluino and neutralino
contributions and the SM contribution set to zero.

\vspace{3cm}

\begin{tabular}{|c|c|c|c|}  \hline
    & $m_{\tilde{q}}=200$ GeV &$m_{\tilde{q}}=500$ GeV &
$m_{\tilde{q}}=1000$ GeV \\
    \hline
    $x$ & \multicolumn{3}{c|}{$\sqrt{|\mathrm{Im}
(\delta^{d}_{12})_{LL}^{2}|},~~~ 
(\delta^{d}_{12})_{RR}=(\delta^{d}_{12})_{LL}$} \\
\hline
0.25& $3.7\times 10^{-5}$

     & $9.7\times 10^{-5}$

     & $2.0\times 10^{-4}$

\\
1.0& $4.3\times 10^{-5}$

    & $1.1\times 10^{-4}$

    & $2.3\times 10^{-4}$

\\
4.0& $6.3\times 10^{-5}$

    & $1.6\times 10^{-4}$

    & $3.3\times 10^{-4}$

\\
\hline
    $x$ & \multicolumn{3}{c|}{$\sqrt{|\mathrm{Im}
(\delta^{d}_{12})_{LR}^{2}|},~~~(\delta^{d}_{12})_{RL}=(\delta^{d}_{12})_{LR} 
$} \\
\hline
0.25& $6.7\times 10^{-5}$

     & $1.6\times 10^{-4}$

     & $3.1\times 10^{-4}$

\\
1.0& $5.5\times 10^{-4}$

    & $7.1\times 10^{-4}$

    & $9.4\times 10^{-4}$

\\
4.0& $1.7\times 10^{-4}$

    & $4.8\times 10^{-4}$

    & $1.1\times 10^{-3}$

\\
\hline
\end{tabular}

\vspace{0.5cm}
${\bf  Table~~ 6} ~~$ Limits on $\mathrm{Im} (\delta_{12}^d)_{AB}
(\delta_{12}^d)_{CD}$,
with A, B, C, D=(L,R) for different values of
$x=m_{\tilde{g}}^2/m_{\tilde{q}}^2$ with two mass insertions each time,
including gluino and neutralino
contributions and the SM contribution set to zero.

\newpage

\begin{tabular}{|c|c|c|c|}  \hline
    $M_L $ $\slash$ $ m_{\tilde{q}}$ (GeV) & 300  & 500 & 700 \\
\hline
        & \multicolumn{3}{c|}{$\sqrt{|\mathrm{Re}
(\delta^{u}_{12})_{LL}^{2}|},~~~ 
(\delta^{u}_{12})_{RR}=(\delta^{u}_{12})_{LL}$}
   \\
\hline
     150 &$4.0 \times 10^{-2}$

         &$5.6 \times 10^{-2}$

         &$7.2 \times 10^{-2}$\\
     250 &$6.8 \times 10^{-3}$

         &$1.0 \times 10^{-2}$

         &$1.5 \times 10^{-2}$\\
     350 &$8.4 \times 10^{-3}$

         &$1.1 \times 10^{-2}$

         &$1.5 \times 10^{-2}$\\
     450 &$8.8 \times 10^{-3}$

         &$1.1 \times 10^{-2}$

         &$1.4 \times 10^{-2}$\\
\hline
         & \multicolumn{3}{c|}{$\sqrt{|\mathrm{Re}
(\delta^{u}_{12})_{LR}^{2}|},~~~(\delta^u_{12})_{RL}=(\delta^u_{12})_{LR} $}
   \\
\hline
     150 &$9.6 \times 10^{-3}$

         &$1.2 \times 10^{-2}$

         &$1.5 \times 10^{-2}$\\
     250 &$7.7 \times 10^{-3}$

         &$1.0 \times 10^{-2}$

         &$1.3 \times 10^{-2}$ \\
     350 &$9.8 \times 10^{-3}$

         &$1.2 \times 10^{-2}$

         &$1.5 \times 10^{-2}$ \\
     450 &$1.1 \times 10^{-2}$

         &$1.3 \times 10^{-2}$

         &$1.6 \times 10^{-2}$ \\
\hline
\end{tabular}

\vspace{0.5cm}
${\bf  Table ~~7} ~~$ Limits on $\mathrm{Re} (\delta_{12}^u)_{AB}
(\delta_{12}^u)_{CD}$,
with A, B, C, D=(L,R) for different values of
$M_L=M_R$ and $m_{\tilde{q}}$ with two mass insertions each time,
with $\tan \beta=5$ and $\mu_H=200$ GeV. The bounds
are insensitive to $\tan \beta$ in the range of $2-30$ and to $\mu_H$
in the range of $200-500$ GeV.

\vspace{ 2 cm}

\begin{tabular}{|c|c|c|c|}  \hline
    $M_L $ $\slash$ $ m_{\tilde{q}}$ (GeV) & 300  & 500 & 700 \\
\hline
        & \multicolumn{3}{c|}{$\sqrt{|\mathrm{Im}
(\delta^{u}_{12})_{LL}^{2}|},~~~ 
(\delta^{u}_{12})_{RR}=(\delta^{u}_{12})_{LL}$}
   \\
\hline
     150 &$3.2 \times 10^{-3}$

         &$4.5 \times 10^{-3}$

         &$5.7 \times 10^{-3}$\\
     250 &$5.5 \times 10^{-4}$

         &$8.3 \times 10^{-4}$

         &$1.2 \times 10^{-3}$\\
     350 &$6.7 \times 10^{-4}$

         &$9.2 \times 10^{-4}$

         &$1.2 \times 10^{-3}$\\
     450 &$7.0 \times 10^{-4}$

         &$9.1 \times 10^{-4}$

         &$1.2 \times 10^{-3}$\\
\hline
         & \multicolumn{3}{c|}{$\sqrt{|\mathrm{Im}
(\delta^{u}_{12})_{LR}^{2}|},~~~(\delta^u_{12})_{RL}=(\delta^u_{12})_{LR} $}
   \\
\hline
     150 &$7.7 \times 10^{-4}$

         &$9.7 \times 10^{-4}$

         &$1.2 \times 10^{-3}$\\
     250 &$6.2 \times 10^{-4}$

         &$8.4 \times 10^{-4}$

         &$1.1 \times 10^{-3}$ \\
     350 &$7.9 \times 10^{-4}$

         &$9.9 \times 10^{-4}$

         &$1.2 \times 10^{-3}$ \\
     450 &$8.7 \times 10^{-4}$

         &$1.1 \times 10^{-3}$

         &$1.3 \times 10^{-3}$ \\
\hline
\end{tabular}

\vspace{0.5cm}
${\bf  Table ~~8} ~~$ Limits on $\mathrm{Im} (\delta_{12}^u)_{AB}
(\delta_{12}^u)_{CD}$,
with A, B, C, D=(L,R) for different values of
$M_L=M_R$ and $m_{\tilde{q}}$ with two mass insertions each time,
with $\tan \beta=5$ and $\mu_H=200$ GeV. The bounds
are insensitive to $\tan \beta$ in the range of $2-30$ and to $\mu_H$
in the range of $200-500$ GeV.

\end{document}